\documentstyle[epsfig,12pt,aps]{revtex}
\renewcommand{\baselinestretch}{1.25}
\newcommand{\be}{\begin{equation}}
\newcommand{\ee}{\end{equation}}
\newcommand{\bee}{\begin{eqnarray}}
\newcommand{\eee}{\end{eqnarray}}

\newcommand{\dj}{d \raisebox{1.4mm}{\hspace*{-2.75mm}-}
                                \hspace*{-1.5mm}}
\begin{document}

\begin{titlepage}
\begin{center}
      {\Large \bf Update of the $\pi N \rightarrow  \eta N$
       and $\eta N \rightarrow \eta N$ 
        partial-wave amplitudes}
\vspace*{10mm}\\
Mijo Batini\'{c}, Ivan  Dadi\'{c}, Ivo \v{S}laus, Alfred \v{S}varc   \\
{\em Ru{\dj}er Bo\v{s}kovi\'{c} Institute,  Zagreb,
Croatia } \\
 B.M.K. Nefkens \\ {\em University of California Los Angeles
     USA} \\ and \\
     T.-S.H. Lee \\ {\em Physics Division, Argonne National 
     Laboratory, Argonne, Illinois, USA}
\end{center}
      A  three-channel, multi-resonance, unitary model developed in 1995
      is       used to determine the $\pi  N  \rightarrow  \eta  N$  and
      $\eta  N  \rightarrow  \eta  N$  amplitudes     using as input the
      latest data for the dominant $S_{11}$ $\pi N$  elastic  scattering
      partial wave following suggestions of Prof.  G.  H\"{o}hler.  The
      sign error in the numerical evaluation of the dispersion  integral
      in the original publication is eliminated.  The remaining weighted
      data set for the $\pi N \rightarrow \eta N$ total and differential
      cross  sections  is  used  as  in  the  original publication.  The
      correction of the numerical error influences  the  $\eta  N$  cusp
      effect  and  improves  the  quality  of the fit to the input data.
      However, our new  result  for  the  $\eta  N$  scattering  length,
      $a_{\eta  N  }=  (0.717 \pm 0.030) + i (0.263 \pm 0.025)$ fm, is a
      sole consequence of the  correction  of  the  $S_{11}$  input  and
      suggests that the $\eta d$ system is unbound or loosely bound.
\end{titlepage}
\renewcommand{\baselinestretch}{1.8}
\newpage
\setcounter{page}{1}
\section{Introduction}
      A  three-channel,  multi-resonance,  unitary  model,  based on the
      formulation developed in Ref.  \cite{Cut79}, has been  applied  in
      Ref.\cite{Bat95a}  to  perform a partial-wave analysis (PWA) using
      the Karlsruhe-Helsinki PWA (KH80) \cite{Hoe83} as  input  for  the
      $\pi   N$   elastic   scattering,   and  the  weighted  total  and
      differential cross section data for the $\pi N \rightarrow \eta N$
      reaction.   The partial-wave amplitudes for the $\pi N \rightarrow
      \eta N$ and $\eta  N  \rightarrow  \eta  N$  transitions  are  the
      predictions  of  the model.  Using the $\eta N$ elastic scattering
      partial-wave  amplitudes  thus  obtained,  the  $\eta  N$   S-wave
      scattering  length  $a_{\eta  N}$  has  been  extracted.   In Ref.
      \cite{Bat95c} it was stressed that the multi-resonance approach is
      essential in determining the value for $a_{\eta N}$.  The $\eta N$
      scattering length reported in that article as well  as  the  value
      given  in  Ref.  \cite{Ari92} are sufficiently large to imply that
      the $\eta d$ bound state  might  exist  as  suggested  by  several
      theoretical predictions\cite{Liu85,Gre96,Wyc96}.

      The purpose of this paper is twofold:  a.  to improve the strongly
      disputed $\pi N$ elastic input in the $S_{11}$ partial  wave  near
      $\eta$  production  threshold, and b.  to correct a numerical sign
      error    \cite{Vra96}    in    the    previous     analyses     in
      Refs.\cite{Bat95a,Bat95c}.   We  shall show that correcting a sign
      error in the  numerical  evaluation  of  the  dispersion  integral
      changes  somewhat  the  predictions  of the model, but the crucial
      parameter for the near  $\eta$  thereshold  calculation  of  other
      processes  -  $\eta  N$  $S$-wave  scattering  length  is strongly
      changed due to the correction of the near $\eta N$ threshold  $\pi
      N$ elastic input only.

      A  sign  error  in  the  numerical  evaluation  of  the dispersion
      integral  Eq.(19)  of   \cite{Bat95a},   see   \cite{Vra96},   has
      influenced the shape and numerical values of all predicted partial
      wave amplitudes.  We give the corrected  values  with  the  dotted
      lines  in  the  Figs.3-10.  The numerical values for the resonance
      parameters and the $\eta N$ $S$-wave scattering length  are  given
      in  Table  1.   We  have  found  that the correction of this error
      influences the shape of the cusp effect, but leads to a better fit
      to  the input $\pi N$ data.  However, as it can be seen from Table
      1.  the $\eta N$ $S$-wave scattering length stays rather high.

      As  a second part of our article we have introduced a new physical
      input by changing the controversial $\pi N$ elastic $S$-wave  near
      $\eta$ production threshold.  The $\pi N$ elastic scattering input
      (KH80) that was used in the analyses of  Refs.\cite{Bat95a,Bat95c}
      has  been critically reviewed recently \cite{Hoe93,Hoe95}.  It was
      concluded that the  S-wave  part  of  the  KH80  solution  can  be
      considerably  improved in the energy range below $\eta$ production
      threshold  \cite{Hoe93,Hoe95}.   It  has  been  shown  within   an
      one-resonance  model  that the suggested modifications in the $\pi
      N$ elastic scattering input introduce about 20 \%  change  in  the
      real   part   of   the   $\eta   N$   S-wave   scattering   length
      \cite{Bat96b,Sva96}.   We   expect   a   similar   change   in   a
      multi-channel,  multi-resonance model.  Since the real part of the
      $\eta N$ S-wave scattering length  is  the  basic  input  for  the
      theoretical  investigations  of  the  possible  existence  of  the
      $\eta$-light nuclei bound states \cite{Liu85,Gre96,Wyc96},  a  new
      fit  with  the  improved $\pi N$ input is clearly needed in making
      progress in  this  direction.   We  have  employed  the  model  of
      Ref.\cite{Bat95a}  to  carry out a new partial-wave analysis using
      the   improved   $\pi   N$   input,    as    suggested    by    G.
      H\"{o}hler\cite{Hoe96}.

      With  the correct treatment of the dispersion integral and the use
      of the improved $\pi  N$  input,  our  new  partial-wave  analysis
      yields  $a_{\eta N} = (0.717 \pm 0.030) + i (0.263 \pm 0.025)$ for
      the $\eta N$ S-wave scattering length.  Its real part is close  to
      the  lowest  value,  Re($a_{\eta N})$=0.7, for the existence of an
      $\eta d$ bound  state  as  predicted  in  Refs.\cite{Gre96,Wyc96}.
      Therefore,  the probability of producing an $\eta$d bound state by
      using various intermediate energy  nuclear  reactions  is  greatly
      reduced.

      In section II, we briefly  review  the  employed  formalism.   The
      input  data used in this analysis is described in section III.  In
      section IV, we present figures showing the fit  to  the  corrected
      and new $\pi N$ elastic scattering input, as well as the resulting
      amplitudes for all transitions between $\pi N$, $\eta  N$  and  an
      effective  $\pi^2  N$  quasi  two-body  channel.  Accordingly, the
      extracted resonance parameters and the  S-wave  scattering  length
      are  given in Tables 1 and 2.

\section{Formalism}
      The  formalism  used  in this work originated from the old CMU-LBL
      analysis   \cite{Cut79},    and    was    presented    fully    in
      Ref.\cite{Bat95a}.    However,  a  sign  error  was  made  in  the
      numerical  evaluation  of  the  dispersion  integral,  Eq.(19)  of
      Ref.\cite{Bat95a},  such that the imaginary part of the dispersion
      integral was taken incorrectly on the lower half  of  the  Riemann
      sheet.    In  Fig.1,  the  dashed  line  shows  our  old  solution
      \cite{Bat95a}, and the dotted line is the solution obtained  after
      the  sign  error  is corrected.  The data (dot points) of the KH80
      solution near $\eta$ production threshold are also displayed .  As
      can  be  seen  in Fig.1.  the numerical error in the evaluation of
      the dispersion integral has      changed the shapes  of  real  and
      imaginary  parts  of the S-wave amplitude.  In addition, the signs
      of the first derivatives of real and imaginary parts also  reverse
      in  the  region  near  the threshold.  The solid curves, which are
      obtained after updating the controversial $\pi N$ elastic $S_{11}$
      input  for  the  energies  up to $\eta$ production threshold, are in
      agreement in shape with the results of the single resonance  model
      of   Bennhold   and   Tanabe   \cite{Ben91}  (dashed  line).   The
      differences in  magnitudes  reflect  the  importance  of  using  a
      multi-channel   and   multi-resonance  approach,  as  stressed  in
      Ref.\cite{Bat95c}.
\section{The new input}
      As we have already mentioned in section I,  there  are  sufficient
      reasons  to  improve  the previously used KH80 solution by using a
      better determined $S_{11}$ partial wave amplitude.  It was pointed
      out  in  Refs.\cite{Hoe93,Hoe95}  that  the  KH80  solution in the
      S-wave is too high when the new data are included.  A complete new
      analysis using the Karlsruhe-Helsinki dispersion relation approach
      has however not been done.  Therefore, a compromising approach  is
      to   use   the   following  recipe  \cite{Hoe96}:   use  the  KH80
      single-energy  solution  everywhere  with  the  exception  of  the
      S$_{11}$  partial wave below 1500 MeV total c.m.  energy where the
      SM95  single-energy  solution  from  Virginia  Polytechnic   group
      \cite{VPI95}  should  be  used.  This is adopted in this work.  We
      have chosen not to compare the full SM95 solution  with  the  KH80
      because they have established quite different number of resonances
      per partial wave.  That, of course, influences the number of  bare
      propagator  initial terms.  In addition, KH80 goes somewhat higher
      in energy then SM95 because of analytical constraints  built  into
      it.   The differences due to this modification are shown in Fig.2.
      The open dots represent the KH80 single-energy solution, while the
      full  dots  represent SM95 single-energy solution.  The transition
      from one solution to the second one is smooth, as can also be seen
      from  the  Fig.2.   Full and dashed curves serve only to guide the
      eye.  The SM95 solution differs significantly from the KH80  below
      1500  MeV c.m.  energy, so distinct consequences upon the $\eta N$
      scattering length in S-wave are expected.  We  mention  here  that
      because of the nature of the model, the fit to other partial waves
      will also be influenced by  this  modification  of  the  input  in
      S-wave.

\section{Results and conclusions}
      The extracted $\eta N$ S-wave scattering length is given in Tables
      1  and  2, together with the extracted resonance parameters in the
      notation of Ref.\cite{Bat95a}, and separately in Fig.3.
      The determined new $\pi N \rightarrow \pi N$, $\pi  N  \rightarrow
      \eta  N$, $\eta N \rightarrow \eta N$ and $\pi N \rightarrow \pi^2
      N$ partial wave amplitudes are the solid  curves  shown  in  Figs.
      4-11.   The old results (dashed curves) from Ref.\cite{Bat95a}, as
      well  as  the  results  after  eliminating  the  afore   described
      numerical  error (dotted lines) are also displayed for comparison.
      We conclude:
      \begin{enumerate}
      \item The $\pi N \rightarrow \eta N$ and $\eta N \rightarrow \eta
      N$ amplitudes for the solution with the KH80 input, but corrected
      for the sign error, are closer to the final solution with the
      modified $\pi N$ elastic $S_{11}$ input. However, the important
      $\eta N$ S-wave scattering  length is drastically changed from
      (0.91 $\pm$ 0.030) + i (0.29 $\pm$ 0.040) to (0.71 $\pm$ 0.030) +
      i (0.263 $\pm$ 0.023) when the uptade of the $S_{11}$ input is
      introduced. That changes the former, erraneous conclusion that the
      $\eta d$ bound state is likely to exist to the correct statement
      that the $\eta d$ bound state is either nonexistent or losely
      bound \cite{Gre96,Wyc96}.
      \item The quality  of  the  overall  fit  to  the  input  data  is
      improved.
          \begin{description}
      \item[a.]  The  fit  (Figs.   4  and 5) to the new $\pi N$ elastic
      scattering input is improved;  particularly for  the  S-wave  (see
      Fig. 4).   It  appears  that  the  correction  of  the error in the
      evaluation   of    the    dispersion    integral,    Eq.(19)    of
      Ref.\cite{Bat95a}, helps the improvement of the fit.
      \item[b.]  The quality of the fit to all of the experimental total
      and differential cross section data for  the  $\pi  N  \rightarrow
      \eta  N$  reaction is practically identical to the fit reported in
      Ref.\cite{Bat95a}.  The large differences in higher partial  waves
      (Figs.   6  and  7)  do not play a significant role in fitting the
      existing data that are mainly in the near threshold energy region.
      The dominant S-wave amplitudes are essentially unchanged.
      \item[c.] There are significant differences between the new (solid
      curves)  and  corrected  old  (dotted  curves)  in  the  $\eta   N
      \rightarrow \eta N$ and $\pi N \rightarrow \pi^2 N$ transitions in
      some partial waves (Figs.8-11).  It is interesting  to  note  that
      the  difference  in  the  $\eta  N$  elastic  scattering in S-wave
      appears to be small, but Fig.  3 shows  that  the  change  in  the
      extracted S-wave scattering length is dramatic.
          \end{description}
      \item  Some  of  the  corrected  and  new values of the resonances
      parameters (see Tables 1 and 2) are different from the old  values
      reported   in   Refs.\cite{Bat95a}.    This  mainly  reflects  the
      sensitivity of a multi-channel, multi-resonance, unitary  approach
      to the input data, as stressed in Refs.\cite{Bat96b}.
      \end{enumerate}
      \newpage
      \noindent
      {\large \bf Acknowledgment} \\
 
      This  work is supported in part by the U.S.  Department of Energy,
      Nuclear Physics Division,  under  contract  No.   W-31-109-ENG-38.
      The  financial  support from Croatia-US grants JF 221 and JF129 is
      also acknowledged.

      \newpage
      \noindent
      {\bf TABLE I.}{\em \ Resonance parameters of the old and corrected
      multiresonance model with 4 $P_{11}$ resonances. }
      The      results      of      elastic     $\pi     N$     analyses
      [1,3,16] are given in the first  column.
      The  results  of  the  partial  wave  analysis  of old publication
      [2] are given  in
      columns  2-6,  and  the  results  of this publication,
      corrected for the sign error,
      are given in
      columns 7-11.  The values  of  the  old  and  corrected  $\eta  N$
      $S$-wave scattering lengths are given at the end.
\begin{center}
\begin{tabular}{ccccccccccc}
\hline\hline
    States       & \multicolumn{5}{c}{Old solution}      &\multicolumn{5}{c}{Corrected solution} \\
                 \cline{2-11}
    L$_{2I,2J}$                  &  Mass        &   Width      & $x_\pi$      & $x_\eta$     & $x_{\pi^{2}}$&  Mass        &   Width      & $x_\pi$      & $x_{\eta} $  & $x_{\pi^2} $   \\
${\rm (_{Mass/Width}^{x_{el}})}$ &  (MeV)       &   (MeV)      & (\%)         &   (\%)       &   (\%)       &  (MeV)       &   (MeV)      &  (\%)        &   (\%)       &   (\%)         \\
\hline \hline
    S$_{11}(_{1535/120}^{38})$   & 1542(6)      &  150(15)     &   34(9)      &   63(7)      &    3(3)      & 1550(9)      &  204(39)     &   39(8)      &   57(7)      &    4(3)        \\
    S$_{11}(_{1650/180}^{61})$   & 1669(17)     &  215(32)     &   94(7)      &    6(5)      &  0.2(2)      & 1659(11)     &  213(20)     &   77(9)      &   13(7)      &   10(4)        \\
    S$_{11}(_{2090/95 \: }^{9})$ & 1713(27)     &  279(54)     &   49(21)     &    2(3)      &   49(19)     & 1792(23)     &  360(49)     &   35(7)      &   19(10)     &   46(10)       \\  \hline
    P$_{11}(_{1440/135}^{51})$   & 1421(18)     &  250(63)     &   56(8)      &    0(0)      &   44(8)      & 1442(17)     &  438(125)    &   62(4)      &    0(0)      &   38(4)        \\
    P$_{11}(_{1710/120}^{12})$   & 1766(34)     &  185(61)     &    8(14)     &   16(10)     &   76(21)     & 1718(16)     &  195(18)     &   28(20)     &    5(7)      &   67(20)       \\
    P$_{11}$                     & 1760(29)     &  109(32)     &   11(25)     &    3(7)      &   86(22)     & 1737(11)     &  159(26)     &   33(29)     &   12(9)      &   55(29)       \\
    P$_{11}(_{2100/200}^{   9})$ & 2203(70)     &  418(171)    &   11(7)      &   86(7)      &    3(4)      & 2136(41)     &  340(86)     &   16(5)      &   83(6)      &    1(1)        \\  \hline
    P$_{13}(_{1720/190}^{14})$   & 1711(26)     &  235(51)     &   18(4)      &  0.2(1)      &   82(4)      & 1722(19)     &  247(29)     &   18(3)      &    2(2)      &   80(4)        \\  \hline
    D$_{13}(_{1520/114}^{54})$   & 1526(18)     &  143(32)     &   46(6)      &  0.1(0.2)    &   54(6)      & 1523(8)      &  133(12)     &   55(5)      &  0.1(0.1)    &   45(5)        \\
    D$_{13}(_{1700/110}^{   8})$ & 1791(46)     &  215(60)     &    4(5)      &   10(6)      &   86(9)      & 1821(23)     &  141(37)     &    9(6)      &   20(5)      &   71(9)        \\
    D$_{13}(_{2080/265}^{   6})$ & 1986(75)     & 1050(225)    &    9(2)      &    7(4)      &   84(3)      & 2047(65)     &  507(122)    &   17(6)      &    8(3)      &   75(6)        \\  \hline
    D$_{15}(_{1675/120}^{38})$   & 1683(19)     &  142(23)     &   31(6)      &  0.1(0.1)    &   69(6)      & 1679(9)      &  152(8)      &   35(4)      &  0.1(0.2)    &   65(4)        \\
    D$_{15}(_{2100/310}^{   7})$ & 2240(65)     &  761(139)    &    8(4)      &  0.1(1)      &   92(4)      & 2216(27)     &  480(16)     &   13(4)      &  0.1(0.3)    &   87(4)        \\  \hline
    F$_{15}(_{1680/128}^{65})$   & 1674(12)     &  126(20)     &   69(4)      &    1(0.4)    &   30(4)      & 1680(7)      &  142(7)      &   67(3)      &  0.2(0.2)    &   33(3)        \\  \hline
    F$_{17}(_{1990/35 \:}^{ 4})$ &    NF        &    NF        &    NF        &    NF        &    NF        & 2256(455)    & 1926(7444)   &    3(6)      &    2(4)      &   95(9)        \\  \hline
    G$_{17}(_{2190/390}^{14})$   & 2198(68)     &  805(140)    &   19(5)      &  0.1(0.3)    &   81(5)      & 2167(89)     &  505(274)    &   14(12)     &  0.2(1)      &   86(12)
\\ \hline \hline
                                 &              &              &              &              &              &              &              &              &              &
\end{tabular}
  \newline
  NF ... \vspace*{1.0cm} not found
\end{center}
   \begin{center}
   \[
   \begin{array}{c}
   {\eta N \ {\rm \bf S-wave \ scattering \ length} }:
       \left\{
             \begin{array}{rl}
   {\rm \bf old}: &   a_{\eta N} = (0.876 \pm 0.047) + i (0.274 \pm 0.039)    \\
   {\rm \bf corrected}: &   a_{\eta N} = (0.910 \pm 0.030) + i (0.290 \pm 0.040)    \\
             \end{array}
         \right.
    \end{array}
                                                            \]
   \end{center}
   \newpage
   \noindent
      {\bf TABLE II.} {\em Resonance parameters of the corrected and new
      multiresonance model with 4 $P_{11}$ resonances.}
      The results of elastic $\pi N$  analyses  [1,3,16]
      are  given  in  the first column.  The results of the corrected partial wave
      analysis of the  publication [2]  are  given  in  columns
      2-6, and the results of this publication  are given in
      columns 7-11.
      The  corrected and new values of the $\eta
      N$ S-wave scattering lengths are given at the end.
  \newline
\begin{center}
\begin{tabular}  {ccccccccccc}
\hline\hline
    States       & \multicolumn{5}{c}{Corrected solution}      &\multicolumn{5}{c}{New solution} \\
                 \cline{2-11}
    L$_{2I,2J}$                  &  Mass        &   Width      & $x_\pi$      & $x_\eta$     & $x_{\pi^{2}}$&  Mass        &   Width      & $x_\pi$      & $x_{\eta} $  & $x_{\pi^2} $   \\
${\rm (_{Mass/Width}^{x_{el}})}$ &  (MeV)       &   (MeV)      & (\%)         &   (\%)       &   (\%)       &  (MeV)       &   (MeV)      &  (\%)        &   (\%)       &   (\%)         \\
\hline \hline
    S$_{11}(_{1535/120}^{38})$   & 1550(9)      &  204(39)     &   39(8)      &   57(7)      &    4(3)      & 1553(8)      &  182(25)     &   46(7)      &   50(7)      &    4(2)        \\
    S$_{11}(_{1650/180}^{61})$   & 1659(11)     &  213(20)     &   77(9)      &   13(7)      &   10(4)      & 1652(9)      &  202(16)     &   79(6)      &   13(5)      &    8(3)        \\
    S$_{11}(_{2090/95 \: }^{9})$ & 1792(23)     &  360(49)     &   35(7)      &   19(10)     &   46(10)     & 1812(25)     &  405(40)     &   32(6)      &   22(10)     &   46(9)        \\  \hline
    P$_{11}(_{1440/135}^{51})$   & 1442(17)     &  438(125)    &   62(4)      &    0(0)      &   38(4)      & 1439(19)     &  437(141)    &   62(4)      &    0(0)      &   38(4)        \\
    P$_{11}(_{1710/120}^{12})$   & 1718(16)     &  195(18)     &   28(20)     &    5(7)      &   67(20)     & 1729(16)     &  180(17)     &   22(24)     &    6(8)      &   72(23)       \\
    P$_{11}$                     & 1737(11)     &  159(26)     &   33(29)     &   12(9)      &   55(29)     & 1740(11)     &  140(25)     &   28(34)     &   12(9)      &   60(35)       \\
    P$_{11}(_{2100/200}^{   9})$ & 2136(41)     &  340(86)     &   16(5)      &   83(6)      &    1(1)      & 2157(42)     &  355(88)     &   16(5)      &   83(5)      &    1(1)        \\  \hline
    P$_{13}(_{1720/190}^{14})$   & 1722(19)     &  247(29)     &   18(3)      &    2(2)      &   80(4)      & 1720(18)     &  244(28)     &   18(3)      &  0.4(1)      &   82(4)        \\  \hline
    D$_{13}(_{1520/114}^{54})$   & 1523(8)      &  133(12)     &   55(5)      &  0.1(0.1)    &   45(5)      & 1522(8)      &  132(11)     &   55(5)      &  0.1(0.1)    &   45(5)        \\
    D$_{13}(_{1700/110}^{   8})$ & 1821(23)     &  141(37)     &    9(6)      &   20(5)      &   71(9)      & 1817(22)     &  134(37)     &    9(6)      &   14(5)      &   77(9)        \\
    D$_{13}(_{2080/265}^{   6})$ & 2047(65)     &  507(122)    &   17(6)      &    8(3)      &   75(6)      & 2048(65)     &  529(128)    &   17(7)      &    8(3)      &   75(7)        \\  \hline
    D$_{15}(_{1675/120}^{38})$   & 1679(9)      &  152(8)      &   35(4)      &  0.1(0.2)    &   65(4)      & 1679(9)      &  152(8)      &   35(4)      &  0.1(0.1)    &   65(4)        \\
    D$_{15}(_{2100/310}^{   7})$ & 2216(27)     &  480(16)     &   13(4)      &  0.1(0.3)    &   87(4)      & 2217(27)     &  481(17)     &   13(4)      &  0.2(1)      &   87(4)        \\  \hline
    F$_{15}(_{1680/128}^{65})$   & 1680(7)      &  142(7)      &   67(3)      &  0.2(0.2)    &   33(3)      & 1680(7)      &  142(7)      &   67(3)      &  0.4(0.2)    &   33(3)        \\  \hline
    F$_{17}(_{1990/35 \:}^{ 4})$ & 2256(455)    & 1926(7444)   &    3(6)      &    2(4)      &   95(9)      & 2262(470)    & 2036(8235)   &    3(6)      &    2(4)      &   95(8)        \\  \hline
    G$_{17}(_{2190/390}^{14})$   & 2167(89)     &  505(274)    &   14(12)     &  0.2(1)      &   86(12)     & 2125(61)     &  381(160)    &   18(12)     &  0.1(0.3)    &   82(12)
\\ \hline \hline
                                 &              &              &              &              &              &              &              &              &              &
\end{tabular}
\end{center}
   \begin{center}
   \[
   \begin{array}{c}
   {\eta N \ {\rm \bf S-wave \ scattering \ length} }:
       \left\{
             \begin{array}{rl}
   {\rm \bf corrected}: &   a_{\eta N} = (0.910 \pm 0.030) + i (0.290 \pm 0.040)    \\
   {\rm \bf new}: &   a_{\eta N} = (0.717 \pm 0.030) + i (0.263  \pm 0.025 )    \\
             \end{array}
         \right.
    \end{array}
                                                            \]
   \end{center}
\newpage
      \begin{figure}
      \label{fig:1}
\centerline{\psfig{figure=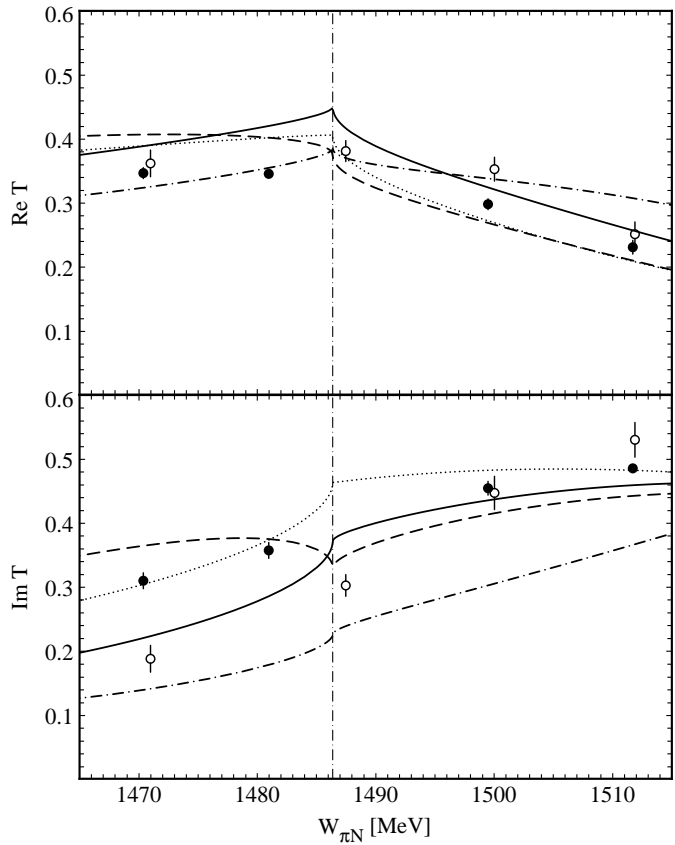,height=12cm,angle=0}}
      \caption{
      Cusp effect in the $\pi N$ elastic $S_{11}$ T-matrix near $\eta N$
      S-wave threshold.  Dashed curves are from  the  solution  obtained
      with  the  wrong  sign  in  numerical  evaluation  of  Eq.(19)  of
      \protect{\cite{Bat95a}}.  Dotted  curves  are  from  the  solution
      corrected  for  the numerical error in the dispersion integral and
      full curves are from the new solutions (solutions  using  modified
      $S_{11}$  $\pi N$ elastic input).  The full dots are from the KH80
      single-energy solution near threshold, and the full ones are  from
      SM95  VPI  solution.  The dash-dotted curves are from the single -
      resonance model by Bennhold and Tanabe \protect{\cite{Ben91}}  for
      comparison.
                                         }
      \end{figure}
      \begin{figure}
      \label{fig:2}
\centerline{\psfig{figure=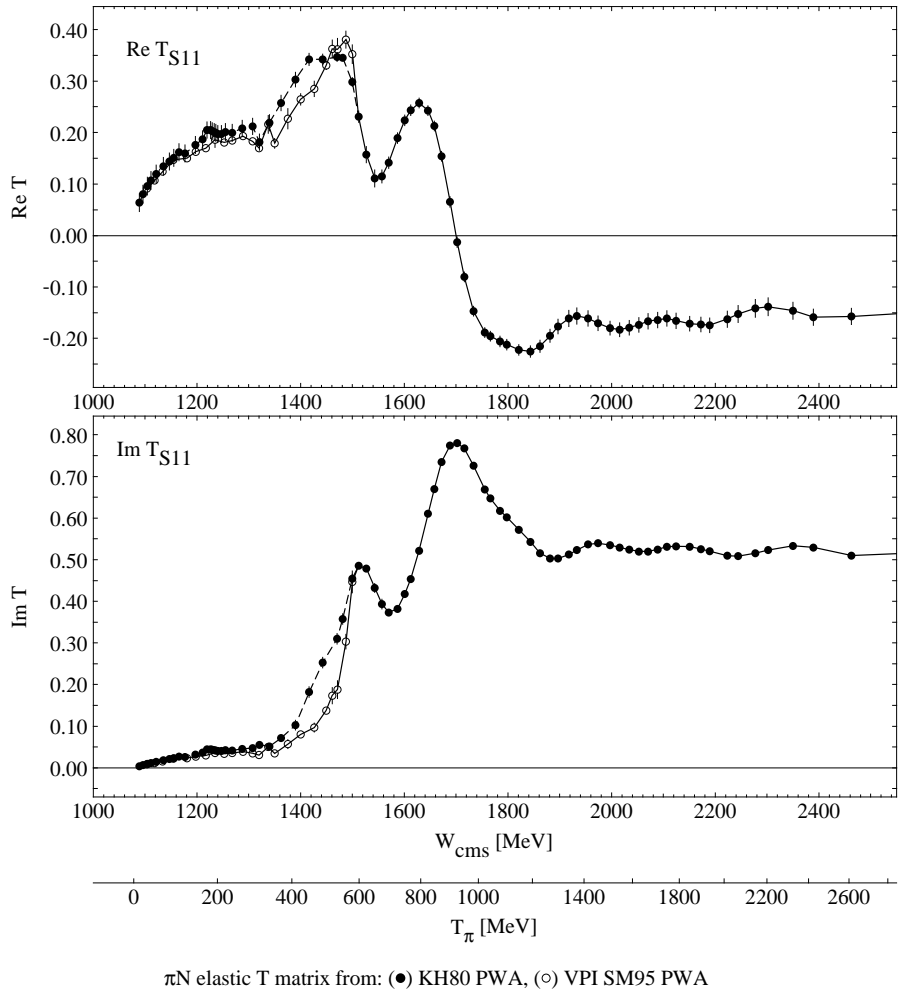,height=14cm,angle=0}}
      \caption{
      The new $\pi  N$  elastic  scattering  input.   The  full  circles
      represent  the old KH80 solution.  The open circles (introduced in
      the S-wave only) represent the replacement of  the  single  energy
      KH80  solution  with  the  single  energy SM95 solution.  Full and
      dashed lines serve only to guide the eye.
                                         }
      \end{figure}
      \begin{figure}
      \label{fig:3}
\centerline{\psfig{figure=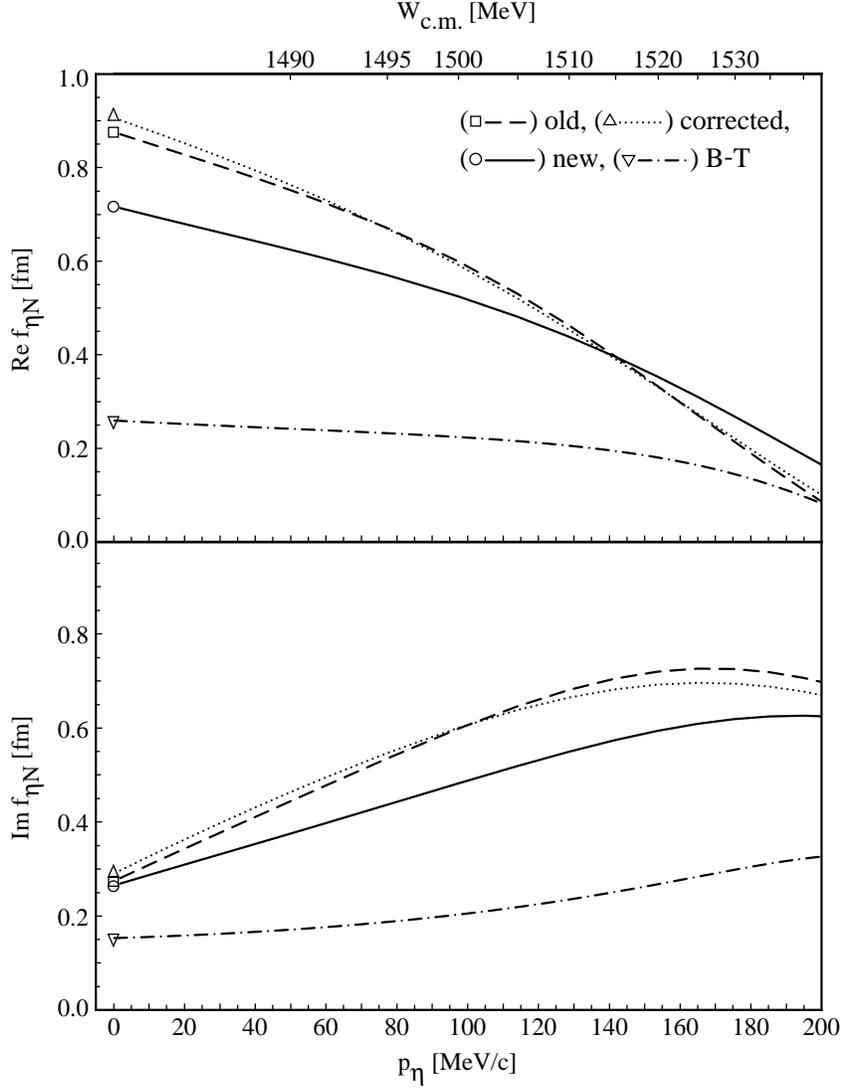,height=18cm,angle=0}}
      \caption{
      The dependence of the $\eta N$ S-wave scattering amplitude defined
      as $f_{0}(p)=T_{0}^{\eta N \rightarrow \eta N} (p)/p_{\eta}$  upon
      the $\eta$ momentum.  The lines have the same meaning as in Fig.1.
                                         }
      \end{figure}

      \begin{figure}
      \label{fig:4}
      \centerline{\psfig{figure=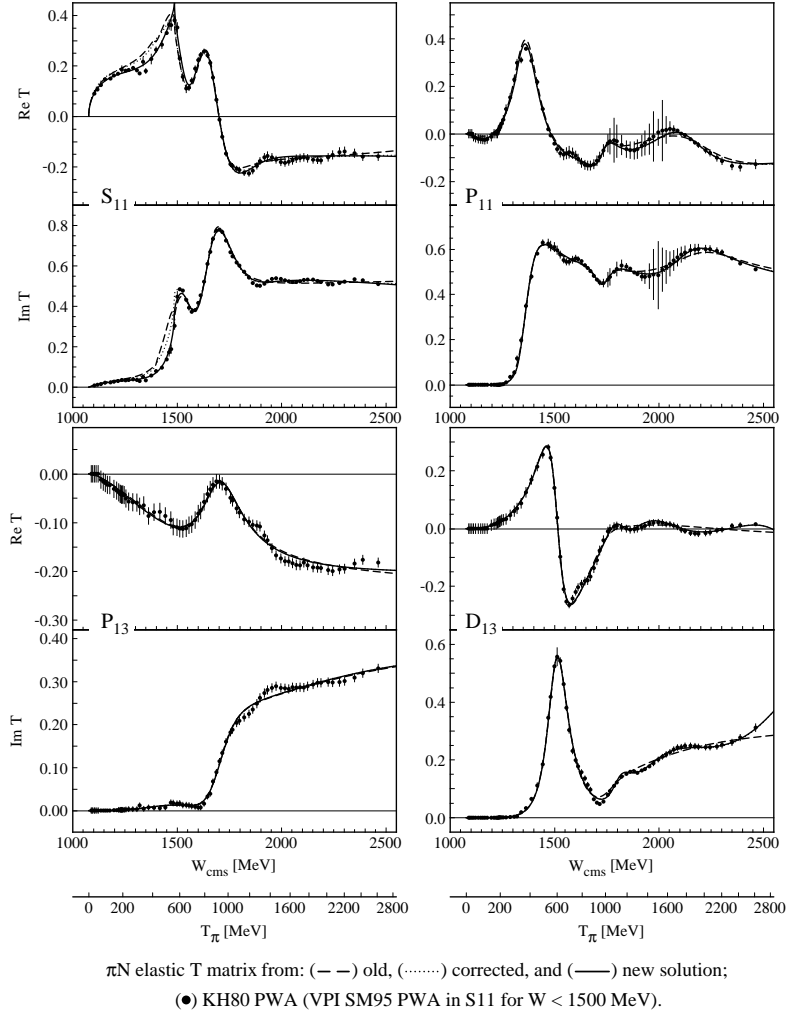,height=14cm,angle=0}}
      \caption{
      The  $\pi  N$ elastic scattering amplitudes in S$_{11}$, P$_{11}$,
      P$_{13}$ and D$_{13}$ partial waves.  The filled circles  are  the
      single-energy  $\pi N$ elastic scattering amplitudes given in Ref.
      \protect{\cite{Hoe83}}      combined      with      the       SM95
      \protect{\cite{VPI95}} for the S-wave below 1500 MeV c.m.  energy.
      The dashed curves are the result  of  the  three  coupled  channel
      multiresonance  model  of  Ref.   \protect{\cite{Bat95a}} with the
      number of resonances given by the PDG \protect{\cite{Pdg92}}.  The
      dotted  line  is  the  result of the same model with the corrected
      numerical error in the evaluation of the dispersion integral.  The
      full  line  is  the  result  of the same model, but with the afore
      described modification of the $\pi N$ elastic $S_{11}$  T-matrices
      data.  }
      \end{figure}

      \begin{figure}
      \label{fig:5}
\centerline{\psfig{figure=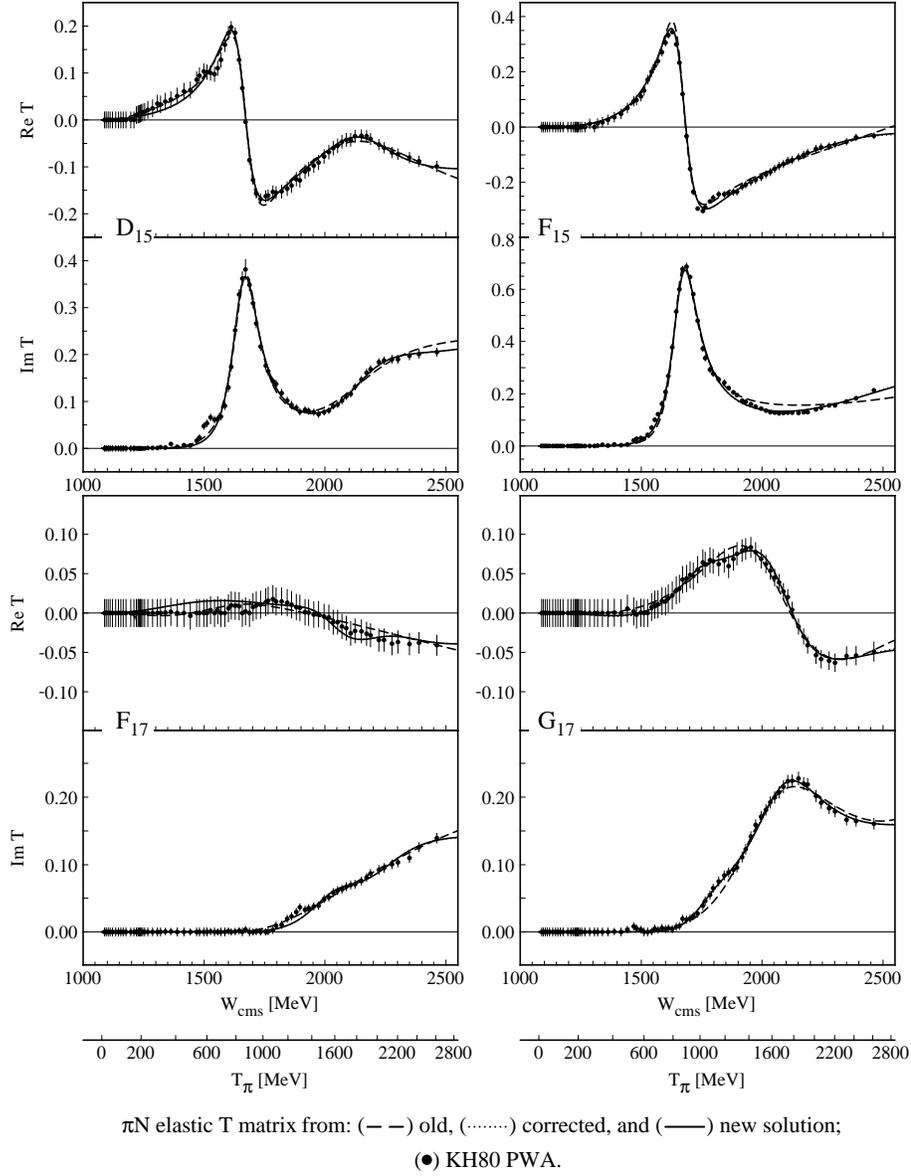,height=16cm,angle=0}}
      \caption{
      The $\pi N$ elastic  scattering amplitudes in 
      D$_{15}$,  F$_{15}$,  F$_{17}$  and  G$_{17}$
      partial  waves.   The filled circles are the single energy $\pi N$
      elastic PWA given in Ref.  \protect{\cite{Hoe83}}.
      The meaning of the different curves is given in the caption of Fig.4.
                                 }
      \end{figure}

      \begin{figure}
      \label{fig:6}
\centerline{\psfig{figure=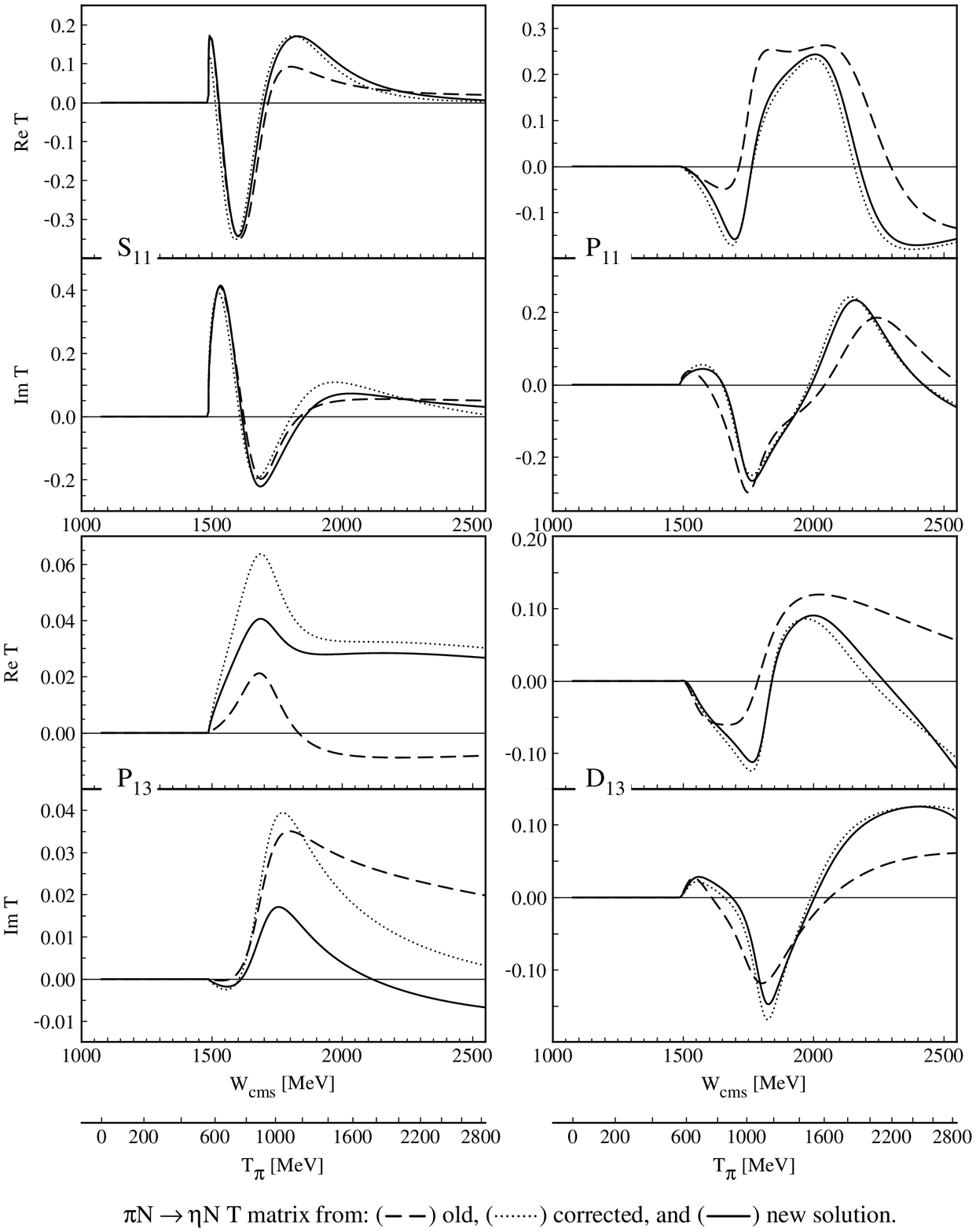,height=16cm,angle=0}}
      \caption{
      The $\pi N \rightarrow \eta N$ amplitudes in 
      S$_{11}$, P$_{11}$,  P$_{13}$  and  D$_{13}$
      partial waves.
      The meaning of the different curves is given in the caption of Fig.4.
                                 }
      \end{figure}

      \begin{figure}
      \label{fig:7}
\centerline{\psfig{figure=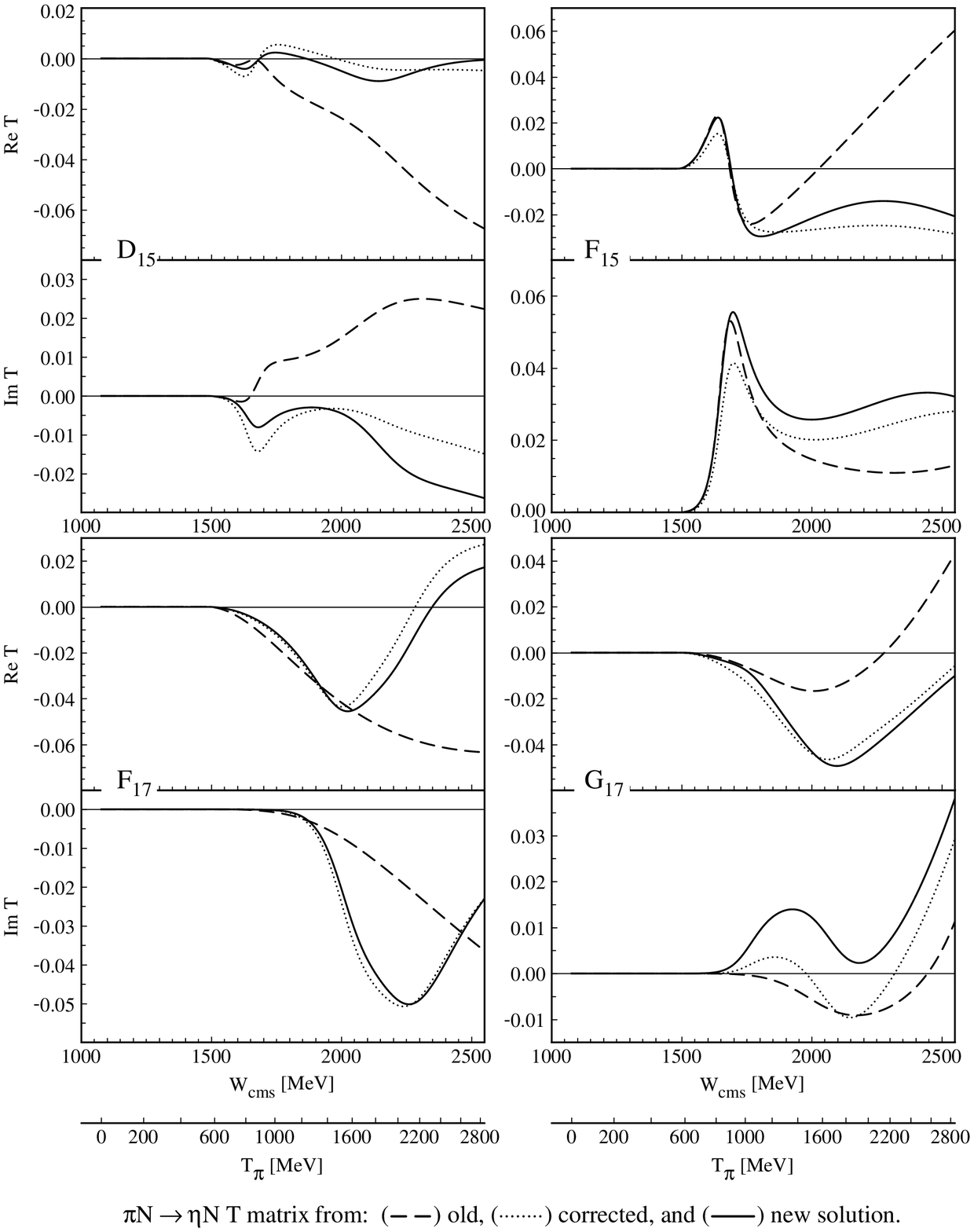,height=16cm,angle=0}}
      \caption{
      The $\pi N \rightarrow \eta N$ amplitudes in
      D$_{15}$, F$_{15}$,  F$_{17}$  and  G$_{17}$
      partial waves.
      The meaning of the different curves is given in the caption of Fig.4.
                                                      }
      \end{figure}

      \begin{figure}
      \label{fig:8}
\centerline{\psfig{figure=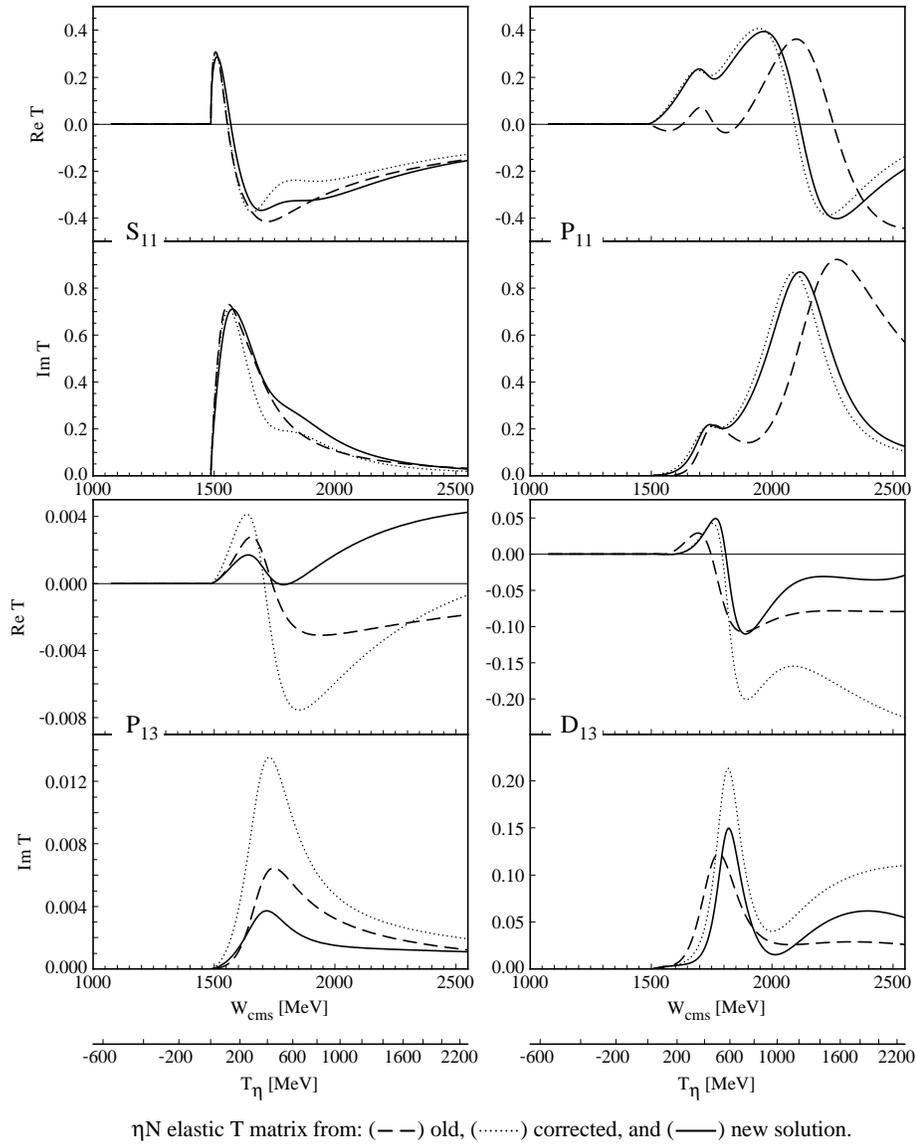,height=16cm,angle=0}}
      \caption{
      The $\eta N$  elastic  scattering amplitudes in 
      S$_{11}$,  P$_{11}$,  P$_{13}$  and  D$_{13}$
      partial waves.
      The meaning of the different curves is given in the caption of Fig.4.
                                 }
      \end{figure}

      \begin{figure}
      \label{fig:9}
\centerline{\psfig{figure=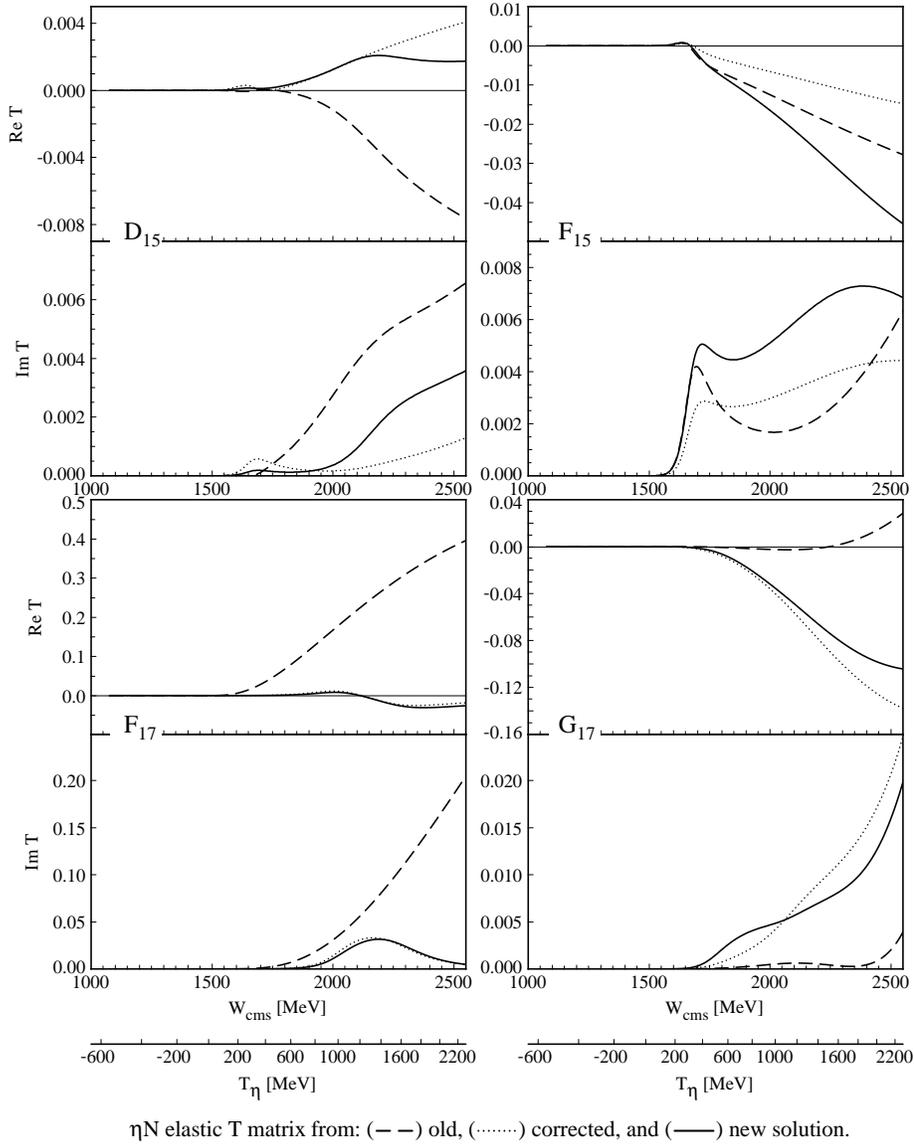,height=16cm,angle=0}}
      \caption{
      The $\eta N$  elastic scattering amplitudes in 
      D$_{15}$,  F$_{15}$,  F$_{17}$  and  G$_{17}$
      partial waves.
      The meaning of the different curves is given in the caption of Fig.4.
                                 }
      \end{figure}

      \begin{figure}
      \label{fig:10}
\centerline{\psfig{figure=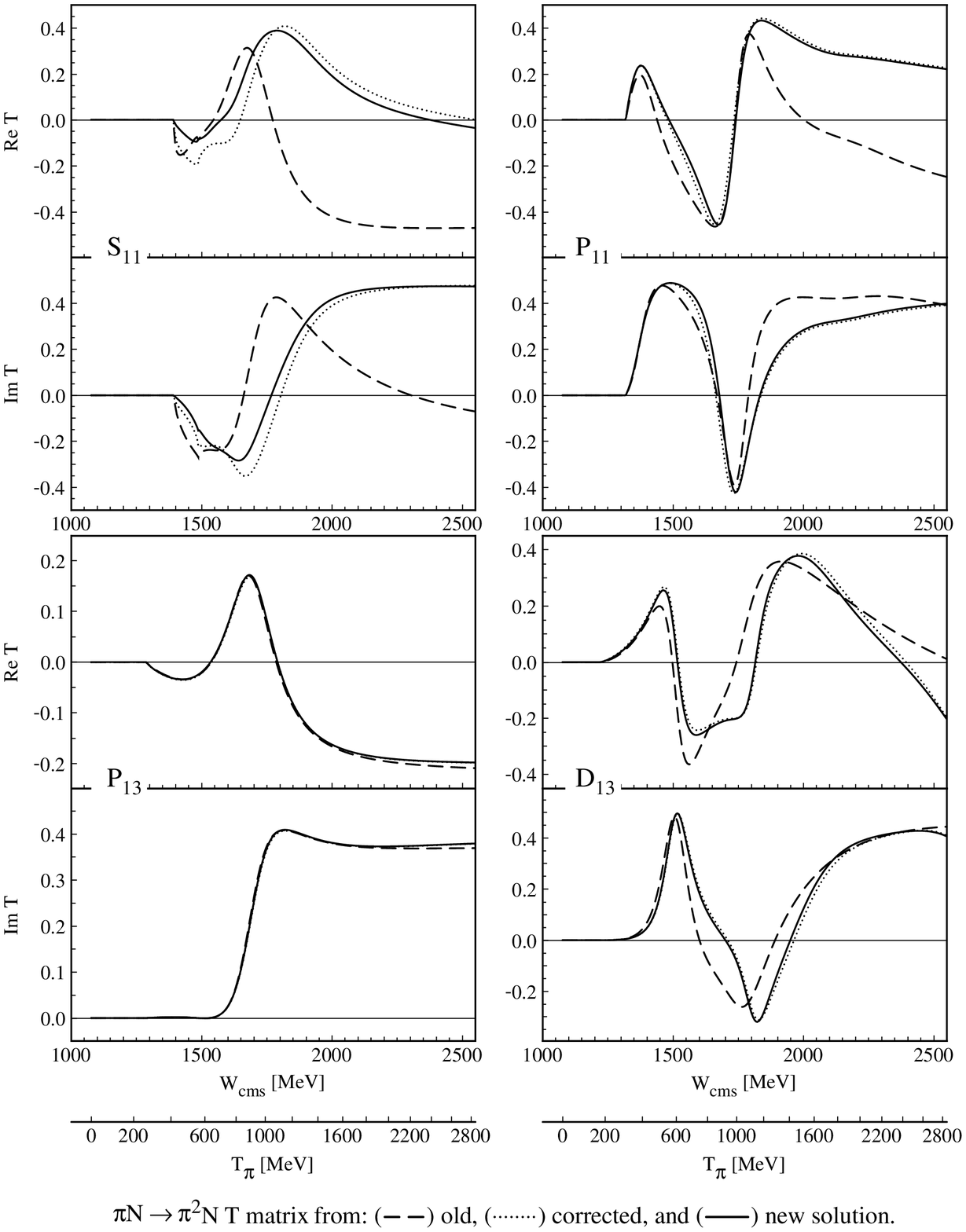,height=16cm,angle=0}}
      \caption{
      The $\pi N \rightarrow \pi^2 N$ amplitudes in 
      S$_{11}$, P$_{11}$,  P$_{13}$  and
      D$_{13}$  partial  waves.
      The meaning of the different curves is given in the caption of Fig.4.
                                 }
      \end{figure}

      \begin{figure}
      \label{fig:11}
\centerline{\psfig{figure=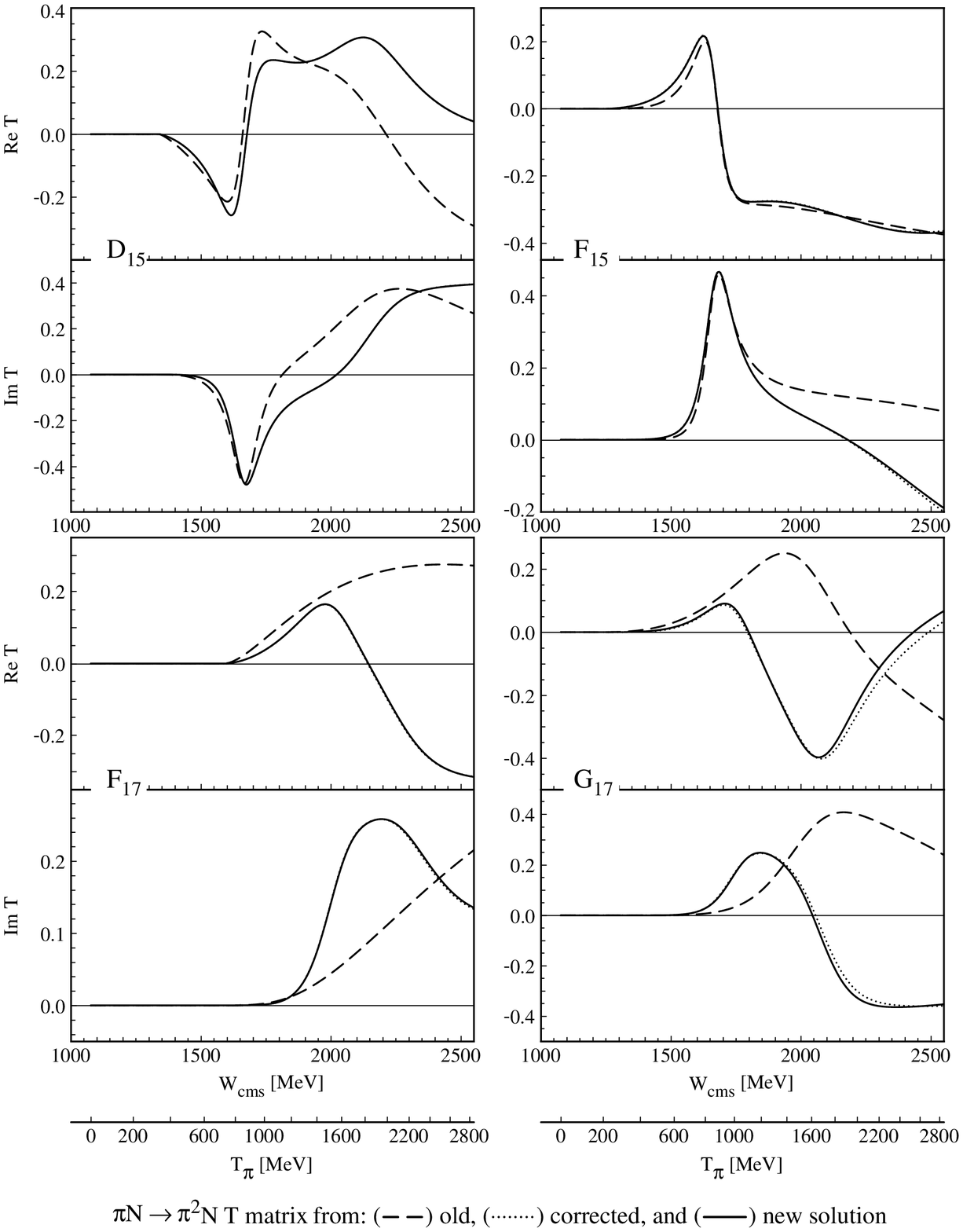,height=16cm,angle=0}}
      \caption{
      The  $\pi  N \rightarrow \pi^2 N$ amplitudes in
      D$_{15}$, F$_{15}$, F$_{17}$ and
      G$_{17}$ partial waves.
      The meaning of the different curves is given in the caption of Fig.4.
                                 }
      \end{figure}

\end{document}